\documentclass[12pt]{article}

\usepackage{amsmath,amsfonts}
\usepackage{amsthm}
\usepackage{graphicx}

\newcommand{\1}{\mathbb{I}}
\newcommand{\btemp}{{B}}
\newcommand{\comap}{\mathcal{L}}
\newcommand{\gogaf}{{G}}
\newcommand{\gogag}{{F}}
\newcommand{\h}{{H}}
\newcommand{\hh}{\mathbf{H}}
\newcommand{\husic}{\chi}
\newcommand{\husrep}{\mathcal{M}}
\newcommand{\intsep}{{\mathcal D}}
\newcommand{\M}{k}
\newcommand{\N}{N}
\newcommand{\pproj}{{\rm P}}
\newcommand{\rfield}{\mathbb{R}}
\newcommand{\rmd}{{\rm d}}
\newcommand{\rme}{{\rm e}}
\newcommand{\rmi}{{\rm i}}
\newcommand{\rrh}{{\rho}}
\newcommand{\sepv}{{\mathbf P}}
\newcommand{\sadj}{\mathcal{E}}
\newcommand{\csp}{\mathcal{N}}

\newtheorem{theorem}{Theorem}

\newtheorem{proposition}[theorem]{Proposition}

\newcommand{\bra}[1]{\pmb{\langle}#1\pmb{|}}
\newcommand{\ket}[1]{\pmb{|}#1\pmb{\rangle}}
\newcommand{\bracket}[3]{\bra{#1}{#2}\ket{#3}}

\newcommand{\gogam}{\gogaf_{\M}}

\newcommand{\mesn}{{\rmd \sigma({\pproj}})}

\DeclareMathOperator{\trc}{Tr}

\title{Husimi coordinates of multipartite separable states}

\author{
Georges Parfionov and Rom\`an R. Zapatrin\\
\small\itshape
Informatics Dept., The State Russian Museum,\\
\small\itshape
In\.zenernaya 4, 191186, St.Petersburg, Russia;\\
\small\rm e-mail: Roman.Zapatrin at Gmail.com }

\begin{document}

\maketitle

\begin{abstract}
A parametrization of multipartite separable states in a
finite-dimensional Hilbert space is suggested. It is proved to be
a diffeomorphism between the set of zero-trace operators and the
interior of the set of separable density operators. The result is
applicable to any tensor product decomposition of the state space.
An analytical criterion for separability of density operators is
established in terms of the boundedness of a sequence of
operators.
\end{abstract}

\section*{Introduction}

We explore multipartite quantum systems with state space
$\h^{(1)}\otimes\cdots\otimes\h^{(\N)}$, assuming the factors are
Hilbert spaces of \emph{finite} dimension:
\[
\dim \h^{(1)},\ldots,\dim \h^{(\N)}<\infty
\]
A pure state of the system is a state whose density operator is
one-dimensional projector. Among pure states we there states of
special kind:
\[
\pproj = P^{(1)}\otimes\cdots\otimes P^{(\N)}
\]
where $P^{(k)}$ is a one-dimensional projector in $\h^{(k)}$. A
density operator $\rrh$ representable as a convex combination
\[
\rrh=\sum w_{i} \pproj_{i},\qquad \pproj_{i}\in \sepv
\]
where $\sepv$ is the set of all pure product states, is said to be
separable. Such description of the set of separable density
operators is not constructive as their appropriate representation
is \emph{not unique}. In this paper a uniform representation of
separable density operators $\rrh$ in the form
\[
    \rrh
    \;=\;\frac{1}{Z}
    \int_\sepv\limits
    \rme^{\trc(\hat{\beta}\cdot\pproj)}
    \pproj\mesn\]
is suggested. This representations turns out to be \emph{unique},
therefore, it can be treated as coordinatization of the set of
separable density operators. Along these lines, a criterion to
determine the separability of a given density operator $\rrh$ is
formulated.

The paper is organized as follows. In Section \ref{shusimi}
separable density operators are treated as barycenters of
\emph{continuously} distributed unit masses on the set $\sepv$ of
pure product states. This gives rise to the equation
\eqref{emainexp} on which the coordinatization of separable
density operators is based. In Section \ref{sconv} a scalar
function \eqref{egogaf} is studied discriminating the separability
in many-particle setting. This makes it possible to formulate a
criterion of multipartite separability in terms of boundedness of
certain sequence of operators. In Section \ref{shusico} the range
of the suggested coordinatization of separable density operators
is determined. The range turns out to be the interior of the set
of separable density operators, and the coordinatization is proved
to be a diffeomorphism.


\section{Separable density operators viewed as barycenters}\label{shusimi}

In this section we describe how density operators in a
finite-dimensional Hilbert space can be described as barycenters
of continuous probability distributions on a set of
one-dimensional projectors.

\subsection{From finite sums to continuous distributions}
We study a multipartite quantum system whose state space $\hh$ is
a finite tensor product of finite-dimensional Hilbert spaces \(
\hh= \h^{(1)}\otimes\cdots\otimes\h^{(\N)} \). A density operator
$\rrh$ in $\hh$ is called \emph{separable} whenever it is an
element of the convex hull of the set $\sepv$ of pure product
states. Since $\sepv$ is the set of extreme points of the set of
all separable density operators, Choquet theorem makes it possible
to write down any such $\rrh$ as an integral
\begin{equation}\label{edefsep}
    \rrh
    =
    \int_\sepv
    \pproj
    \rmd\mu(\pproj)
\end{equation}
over certain probability measure $\mu$ on $\sepv$. In other words,
we may represent $\rrh$ as the \emph{barycenter} of certain
distribution of unit mass on the set $\sepv$. Such representation
is known to be essentially \emph{non-unique}. According to
Carath\'eodory theorem this mass may have discrete finite
distribution
\[
\rrh=\sum w_{\alpha} \pproj_{\alpha}
\]
with the measure $\mu=\sum w_{\alpha} \delta_{\pproj_{\alpha}}$
being a sum of atomic measures $\delta_{\pproj_{\alpha}}$.

\medskip

We emphasize that representation \eqref{edefsep} also comprises
\emph{continuous} distributions
\begin{equation}\label{egenrepint}
    \rrh
    \;=\;
    \int_\sepv
    \pproj w(\pproj)
    \mesn
\end{equation}
where $w(\pproj)$ is a positive continuous function and the
integration is carried out over the probability measure
$\rmd\sigma$ \emph{invariant} with respect to all \emph{local}
unitary transformations in $\hh$.

So, a density operator $\rrh$ in the product space \( \hh=
\h^{(1)}\otimes\cdots\otimes\h^{(\N)} \) is separable when it is a
barycenter of a continuous distribution $w(\pproj)$ on the set
$\sepv$ of pure product vectors. In other words, that means that a
function $w(\pproj)$ exists satisfying the equation
\begin{equation}\label{emaineq}
    \int_\sepv
    \pproj w(\pproj)
    \mesn
    \;-\;
    \rrh
    \;=\;0
\end{equation}

\subsection{Main equation}\label{smaineq}
In this section we confine ourselves to distributions $w(\pproj)$
of specific form and explore the existence of appropriate
solutions.

\medskip

Let us first try to find solutions of \eqref{emaineq} of the form
\begin{equation}\label{egexp}
    w(\pproj)
    =
    \rme^{\trc(\btemp\pproj)}
\end{equation}
where $\btemp$ is an Hermitian operator. Then solving the equation
\eqref{emaineq} reduces to finding an operator $X$ satisfying what
we call \emph{main equation}
\begin{equation}\label{emainexp}
    \int_\sepv
    \pproj
    \rme^{\trc(X\pproj)}
    \mesn
    \;-\;
    \rrh
    \;=\;0
\end{equation}
In order to verify the existence of a solution of
\eqref{emainexp}, introduce the function
\begin{equation}\label{egogaf}
    \gogaf(X)
    \;=\;
    \int_\sepv
    \rme^{\trc(X\pproj)}
    \mesn
    \;-\;
    \trc(X\!\rrh)
\end{equation}
whose gradient is
\begin{equation}\label{egrad}
    \nabla \gogaf(X)
    \;=\;
    \int_\sepv
    \pproj
    \rme^{\trc(X\pproj)}
    \mesn
    \;-\;
    \rrh
\end{equation}
So, solving \eqref{emainexp} reduces to finding an extremum of the
function $\gogaf(X)$. Taking into account that $\gogaf(X)$ is
\emph{convex}, the solution of \eqref{emainexp} exists only when
the \emph{minimum} of $\gogaf(X)$ exists. In particular, when
$\rrh$ is entangled, there is no way for the function $\gogaf(X)$
to have minimum:
\begin{proposition}\label{tentwitn} If the density operator $\rrh$ is
entangled, then
\[
\inf\gogaf(X) \;=\; -\infty
\]
\end{proposition}
\begin{proof}
The set of all separable states is closed, therefore, if $\rrh$ is
not separable, there exists a hyperplane, defined by a
self-adjoint operator $X$ such that $\forall
\pproj\in\sepv\quad\trc(\pproj X)<0$, while $\trc(\rrh \,X)>0$.
Denote
\[
\begin{array}{rcl}
  a&=&\trc(\rrh \,X)\\
  b&=&\max\trc(\pproj \,X)
\end{array}
\]
Then $a>0, b<0$, so $\rme^{\trc(X\pproj)}\leq\rme^{k b}$ and
\[
\gogaf(k X) \;\leq\; \int \rme^{k b}\mesn- k a \;\to\;-\infty
\]
as $k\to\infty$.
\end{proof}

So far, finding conditions for the existence of minimum becomes
essential to judge if $\rrh$ is entangled or not.

\section{Exponential distributions and their
approximations}\label{sconv}

\subsection{An interlude on minima of convex
functions}\label{stech}

\begin{proposition}\label{tminrays}
Let a convex function $\gogag(X)$ has minimum on a
finite-dimensional space $\sadj$ and the minimal point is unique.
Then
\begin{equation}\label{einftymin}
\lim_{X\to\infty}\gogag(X)=+\infty
\end{equation}
\end{proposition}
\begin{proof}With no loss of generality assume the minimum to be
attained at 0, then $\gogag(X)>\gogag(0)$ for any $X\neq 0$.
Suppose $\lim_{X\to\infty}\gogag(X)\neq +\infty$, then there
exists such $M$ and such sequence $X_{k}\to\infty$ that
$\gogag(X_{k})\leq M$. Then
\[
\lambda_{k}=\frac{1}{\Vert X_{k}\Vert}\to 0
\]
Denoting
\[
E_{k} \;=\; \frac{X_{k}}{\Vert X_{k}\Vert}\;=\;\lambda_{k} X_{k}
\]
we obtain a bounded sequence in the finite-dimensional space
$\sadj$, which contains a converging subsequence. With no loss of
generality denote this subsequence $E_{k}$ and its limit \[ E=\lim
E_{k}, \quad \Vert E \Vert = 1\]
Then
\[
E_{k} \;=\;\lambda_{k} X_{k}\;=\;(1-\lambda_{k})\,0+\lambda_{k}
X_{k}
\]
so, due to the convexity of $\gogag$
\[
\gogag(E_{k})\leq (1-\lambda_{k})\,\gogag(0)+\lambda_{k}
\gogag(X_{k})
\]
therefore
\[
\gogag(E_{k})\leq (1-\lambda_{k})\,\gogag(0)+\lambda_{k} M
\]
for sufficiently large $k$. Taking the limit and recalling that
$\lambda_{k}\to 0$, and using the continuity of convex functions
(see, e.g. \cite{convexcont}), we come to the contradiction:
$\gogag(E)\leq\gogag(0)$.
\end{proof}

So far, we obtained a necessary condition for the unique minimum
of a convex function to exist. This condition \eqref{einftymin} is
also sufficient for a minimum (not necessarily strict) to exit. In
the sequel we shall need a more verifiable sufficient condition:

\begin{proposition}\label{tinftyrays}
Let $\gogag(X)$ be a convex function on a finite-dimensional space
$\sadj$. If
\begin{equation}\label{erays}
    \forall\,X\neq0 \qquad \lim_{t\to\infty}\gogag(tX)=+\infty
\end{equation}
then
there exists minimum of $\gogag(X)$ on $\sadj$.
\end{proposition}
\begin{proof}
First prove that \[\lim_{X\to\infty}\gogag(X)=+\infty\] Suppose
this is not the case, then there exists a number $M$ and a
sequence $X_{k}\to\infty$ such that $\gogag(X_{k})\leq M$ for all
$k$. Denoting
\[
E_{k} \;=\; \frac{X_{k}}{\Vert X_{k}\Vert}
\]
we obtain a bounded sequence in the finite-dimensional space
$\sadj$, which contains a converging subsequence. With no loss of
generality denote this subsequence $E_{k}$ and its limit \[ E=\lim
E_{k}, \quad \Vert E \Vert = 1\] Now take an arbitrary $t\geq 0$
and $N$ such that $\Vert X_{k}\Vert\geq t$ for all $k\geq N$. Then
$\Vert t E_{k}\Vert=t\leq\Vert X_{k}\Vert$. Denote
\[
\lambda_{k}=\frac{t}{\Vert X_{k}\Vert}
\]
then $0\leq \lambda_{k}\leq 1$ and $t E_{k}=\lambda_{k} X_{k}$. We
may treat $t E_{k}$ as convex combination
\[
t E_{k}=(1-\lambda_{k})\,0+\lambda_{k} X_{k}
\]
and apply Jensen's inequality
\[
\gogag(t E_{k}) \leq (1-\lambda_{k})\gogag(0)+\lambda_{k}
\gogag(X_{k}) \leq \gogag(0) + M
\]
for all $k\geq N$. Taking the limit, we get $\gogag(t E)\leq
\gogag(0) + M$, which contradicts with the statement $\gogag(t
E)\to\infty$.

Now let us prove the existence of minimum. For that, consider the
set
\[
K=\{X\mid \gogag(X)\leq \gogag(0)\}
\]
The function $\gogag$ is convex (and hence continuous), so the set
$K$ is closed. As proved above,
$\lim_{X\to\infty}\gogag(X)=+\infty$, therefore the set $K$ is
bounded.

The space $\sadj$ is finite-dimensional, the set $K$ is compact,
therefore the function $\gogag$ always attains minimum on $K$,
which is its minimum on the whole space $\sadj$.
\end{proof}

So far, we have obtained a sufficient condition for the existence
of the minimum of a convex function.

\subsection{Multipartite setting}\label{smult}

Now return to our initial setting. We are dealing with integrals
over the set $\sepv$ of pure product projectors in a tensor
product space $\hh= \h^{(1)}\otimes\cdots\otimes\h^{(\N)}$. We
shall need the following:
\begin{proposition}[\textbf{Multipolarization identity}]\label{tpolariz}
Any quadratic form $Y$ in $\hh$ is completely defined by its
values on $\sepv$.
\[
    \bracket{e^{(1)}\otimes\cdots\otimes
e^{(\N)}}{Y}{f^{(1)}\otimes\cdots\otimes
f^{(\N)}}=\]
\begin{equation}\label{epolariz} =4^{-\N}\!\!\!\!
\sum_{k^{(1)},\ldots\,k^{(\N)}=0}^3\limits\;
\!\!\rmi^{(k^{(1)}+\cdots+k^{(\N)})}
\left\langle\bigotimes_{I=0}^{\N}\limits
\left(e^{(I)}+\rmi^{k^{(I)}}f^{(I)}\right)\right.
\left|\vphantom{\bigotimes}{Y}\right| \bigotimes_{I=0}^{\N}\limits
\left.\left(e^{(I)}+\rmi^{k^{(I)}}f^{(I)}\right)
\vphantom{\bigotimes_{I=0}^{\N}\limits}\right\rangle
\end{equation}
\end{proposition}
\begin{proof}Verified by direct calculation.\end{proof}

\paragraph{Remark.} In the sequel we shall use the following consequence
of this proposition:
\begin{equation}\label{edegen}
    \forall \pproj\in\sepv\quad \trc(X\pproj)=0\quad\Rightarrow\quad X=0
\end{equation}

Now return to the convex function \eqref{egogaf}
\[
    \gogaf(X)
    \;=\;
    \int_\sepv
    \rme^{\trc(X\pproj)}
    \mesn
    \;-\;
    \trc(X\!\rrh)
\]
\begin{proposition}\label{tgexist}
When the function $\gogaf(X)$ has minimum, this minimum is strict.
\end{proposition}
\begin{proof}
Suppose there are two minimal points $\btemp_0\neq \btemp_1$ of
$\gogaf(X)$. Denote $V=\btemp_1-\btemp_0$ and consider a family
$\btemp_t=\btemp_0+Vt$. Consider the function
\[
g(t) \;=\; \gogaf(\btemp_t)
    \;=\;
    \int_\sepv\limits
    \rme^{\trc(\btemp_t\pproj)}
    \mesn
    \;-\;
    \trc(\btemp_t\!\rrh)
\]
 for $t\in[0,1]$. This function is constant because $\gogaf$ is \emph{convex}.
 Therefore its second derivative vanishes, but:
\begin{equation}\label{ehesse}
g^{\,\prime\prime}(t)
    \;=\;
    \int_\sepv\limits
    \rme^{\trc(\btemp_t\pproj)}\:(\trc(V\pproj))^{2}
    \mesn
\end{equation}
In the meantime $V\neq 0$, therefore $\int_\sepv
    (\trc(V\pproj))^{2}
    \mesn >0$ due to \eqref{edegen}, so $g^{\,\prime\prime}(t)>0$
    --- contradiction.
\end{proof}
However, the \emph{existence} of the minimum of $\gogaf(X)$ still
can not be directly verified. In the meantime, the approximations
of $\gogaf(X)$ by the family of convex functions
\begin{equation}\label{egogam}
    \gogam(X)
    \;=\;
    \int_\sepv
    \left(1+\frac{\trc(X\pproj)}{2\M}\right)^{2\M}
    \mesn
    \;-\;
    \trc(X\!\rrh)
\end{equation}
possess the following property:
\begin{proposition}\label{tgmexist}
Whatever be $\rrh$, for any $\M$ there exists minimum of the
function $\gogam(X)$ and this minimum is strict.
\end{proposition}
\begin{proof}
First prove that the minimum exists. Fix an arbitrary $X\neq 0$
and apply the sufficient condition \eqref{erays} by showing that
$\gogam(tX)\to+\infty$. Let
\[
Y=\1+\frac{t X}{2\M}
\]
then
\[
\gogam(tX) = \int_{\sepv} \trc(YP)^{2\M} \mesn-t\,\trc(XA)
\]
For all $Y\neq 0$, introduce
\[
N(Y) = \frac{\int_{\sepv} \trc(YP)^{2\M}
\mesn}{\trc(Y^2)^{\M}}\geq 0
\]
Being homogeneous, $N(Y)$ is completely defined by its values on
the compact set defined by $\trc(Y^2)=1$. Let us prove that $N(Y)$
is strictly positive. Suppose $N(Y)=0$ for some $Y$, then
$\trc(YP)=0$ for all product one-dimensional projectors
$P\in\sepv$. That means, for all vectors $e^{(1)},\ldots,e^{(\N)}$
\[
\bracket{e^{(1)}\otimes\cdots\otimes
e^{(\N})}{Y}{e^{(1)}\otimes\cdots\otimes e^{(\N)}}=0
\]
Then, by virtue of \eqref{edegen}, $Y=0$.

Being continuous function defined on a compact set, $N(Y)$ attains
its minimal value, denote it $a$. Then
\[
\gogam(tX)\geq a\trc(Y^2)^{\M} -t\,\trc(X\rrh) \;=\; at^{2\M}
\left(\trc\left( t^{-1}\1 + \frac{X}{2\M}
\right)^2\right)^{\M}-t\,\trc(X\rrh) \;\to\;+\infty
\]
since $a>0$ as $N(Y)$ was proved to be strictly positive. So the
minimum of $\gogam(X)$ exists.

In order prove that the minimum is strict, we proceed in a way
similar to Proposition \ref{tgexist} with the only difference that
the function $g(t)$ has the form:
\[
g(t) \;=\; \gogam(\btemp_t)
    \;=\;
    \int_\sepv\limits
    \left(1+\frac{\trc(\btemp_t\pproj)}{2\M}\right)^{2\M}
    \mesn
    \;-\;
    \trc(\btemp_t\!\rrh)
\]
and we check its $2\M$-s derivative (rather than the second one)
\[
g^{(2\M)}(t)
    \;=\;
    \int_\sepv\limits
    \frac{\left({2\M-1}\right)!}{(2\M)^{2\M-1}}\:(\trc(V\pproj))^{2\M}
    \mesn
\]
and obtain the same contradiction.
\end{proof}

\paragraph{Corollary.}  Whatever (separable or entangled) be $\rrh$,
it decomposes into
\begin{equation}\label{egendecomp}
    \rrh
    \;=\;
    \int_\sepv
    \pproj w(\pproj)
    \mesn
\end{equation}
with
\begin{equation}\label{egendens}
w(\pproj)=\left(1+\frac{\trc(\btemp\pproj)}{2\M}\right)^{2\M-1}
\end{equation}
where $\btemp$ is the minimal point of the function $\gogam$, that
is, $\nabla\gogam(\btemp)=0$.

We emphasize that the obtained decomposition \eqref{egendecomp} is
in general \emph{not barycentric} as the density \eqref{egendens}
may take negative values. Furthermore, for entangled $\rrh$ the
density \eqref{egendens} will always take both positive and
negative values. Let us consider it in more detail.

\subsection{The convergence of the approximations}\label{sbinom}

We start from the main equation \eqref{emainexp}
\[
    \int_\sepv
    \pproj
    \rme^{\trc(X\pproj)}
    \mesn
    \;-\;
    \rrh
    \;=\;0
\]
which may not have solutions and replace it by a sequence of its
binomial approximations
\begin{equation}\label{ebinexp}
    \int_\sepv
    \pproj
    \left(1+\frac{\trc(X\pproj)}{2\M}\right)^{2\M-1}
    \mesn
    \;-\;
    \rrh
    \;=\;0
\end{equation}
each always having a unique solution. So, we have to study the
conditions when these approximations turn to the main equation.
For that, we explore the convergence $\gogam\to\gogaf$ of
functions associated with the equations in question on the set
$\sadj$ of self-adjoint operators in $\hh$.
\begin{theorem}\label{t14.7}
If the function $\gogaf(X)$ has strict minimum on $\sadj$ attained
in $\btemp$, then the sequence $\btemp_\M$ of the minimal points
of $\gogam(X)$ converge to $\btemp$.
\end{theorem}
\begin{proof}
First prove that the convergence $\gogam\to\gogaf$ is uniform on
any compact subset of $\sadj$. For any $0\leq\,a\leq n$ direct
calculation yields
\[
\max_{|x|\leq a}\left|e^{x}-\left(1+\frac{x}{n}\right)^n\right|=
\max\big\{e^{a}-\left(1+\frac{a}{n}\right)^n,\;
e^{-a}-\left(1-\frac{a}{n}\right)^n\big\}
\]
therefore
\begin{equation}\label{euniexp}
\left(1+\frac{x}{n}\right)^n - \rme^{x} \to 0
\end{equation}
uniformly on any finite interval in $\rfield$. For $n=2\M$
\[
\left| \gogam(X) - \gogaf(X) \right| \leq \int_{\sepv}\limits
\left| \left(1+\frac{\trc(X\pproj)}{2\M}\right)^{2\M} -
\rme^{\trc(X\pproj)} \right| \mesn
\]
Since $X$ ranges over a compact set, $|\trc(X\pproj)|\leq C$ for
some $C$ not depending on $X$. The integration set $\sepv$ is
compact, so the convergence $\gogam\to\gogaf$ is uniform.

\medskip

Now let us prove that the sequence $\btemp_\M$ tends to $\btemp$.
The minimum is strict, so $\gogaf(X)>\gogaf(\btemp)$ for any
$X\neq \btemp$. The sphere $\|X-\btemp\|=\varepsilon$ is compact,
then there exists
\[
a=\min_{\|X-\btemp\|=\varepsilon}\limits \gogaf(X) -
\gogaf(\btemp)
>0
\]
Since $\gogam\to \gogaf$ uniformly on compact sets, a number
$N_\varepsilon$ exists such that  for any $\M>N_\varepsilon$
\[
\gogaf(X)-\frac{a}{3}\,<\,\gogam(X)\,<\,\gogaf(X)+\frac{a}{3}
\]
for any $X:\:\|X-\btemp\|\leq \varepsilon$. Then for any such $X$
\[
\frac{2}{3} a= a - \frac{a}{3}\leq \gogaf(X)-\gogaf(\btemp) -
\frac{a}{3} < \gogam(X) - \gogaf(\btemp)
\]
So,
\begin{equation}\label{ea10} \frac{2}{3} a+\gogaf(\btemp)< \gogam(X)
\qquad \forall X:\: \|X-\btemp\|\leq \varepsilon
\end{equation}
It remains to check that $\|\btemp_k-\btemp\|\leq \varepsilon$ for
all $k>N_\varepsilon$. Suppose a number $k>N_\varepsilon$ exists
such that the appropriate minimal point
$\|\btemp_k-\btemp\|>\varepsilon$. For
$t=\varepsilon/\|\btemp_k-\btemp\|$ consider the convex
combination
\[
E=(1-t)\cdot \btemp + t \btemp_k
\]
then $\|E-\btemp\|=\varepsilon$. Then it follows from \eqref{ea10}
and the convexity of $\gogam$ that
\[
\frac{2}{3} a + \gogaf(\btemp)< \gogam(E)  \leq (1-t)\cdot \gogam
(\btemp) + t \gogam(B\btemp_k)\leq
\]
\[\leq (1-t)\cdot \gogam (\btemp) + t \gogam(\btemp)= \gogam(\btemp) < \frac{a}{3} +
\gogaf(\btemp)
\]
--- contradiction.

\end{proof}

\begin{theorem}\label{t14.2}
If the sequence $\btemp_\M$ of the minimal points of $\gogam(X)$
converge, then the function $\gogaf(X)$ has strict minimum at
point $\btemp=\lim \btemp_{\M}$.
\end{theorem}
\begin{proof}
For any $X$
\[
\gogam(X)\geq \gogam(\btemp_k)
\]
For any fixed $X$ the sequence $\gogam(X)\to \gogaf(X)$, so
\begin{equation}\label{e14.2}
    \gogaf(X)=\lim \gogam(X)\geq \;\overline{\lim} \;\gogam(\btemp_{\M})
\end{equation}
Now check that
\[
\lim \;\gogam(\btemp_{\M})=\gogaf(\btemp)
\]
The convergence $\btemp_{j}\to \btemp$ and $\gogam\to\gogaf$
together with the continuity of each $\gogam(X)$ imply
\begin{itemize}
 \item $\forall \M\: \gogam(\btemp_{j})\to\gogam(\btemp)$
 \item $\gogam(\btemp)\to\gogaf(\btemp)$
\end{itemize}
Applying Cantor diagonal method, we choose a subsequence
$\btemp_{j_{\M}}$ such that
$\gogam(\btemp_{j_{\M}})\to\gogaf(\btemp)$. For arbitrary $X,Y$ we
have
\[
    |\gogam(X)-\gogam(Y)|
\leq    \int_\sepv\limits
 \left|
    \left(1+\frac{\trc(X\pproj)}{2\M}\right)^{2\M}
    -
    \left(1+\frac{\trc(Y\pproj)}{2\M}\right)^{2\M}
 \right|
    \mesn
    \;+\;\|X-Y\|\cdot\|\rrh\|
\]
Observing that for any $x,y\in\rfield$
\[
 \left|
 \left(
 1+\frac{x}{n}
 \right)^{n}
 -
 \left(
 1+\frac{y}{n}
 \right)^{n}
 \right|
 \leq
 \rme^{|x|+|y|}
 \cdot
 |x-y|
\]
and taking into account that $\trc(X\pproj)\leq \|X\|$, we get
\begin{equation*}\label{e13.5}
    |\gogam(X)-\gogam(Y)|\leq
    \left(
    \rme^{\|X\|+\|Y\|}
    +
    \|\rrh\|
    \right)
    \cdot
    \|X-Y\|
\end{equation*}
Using this, we have
\[
    |\gogam(\btemp_{\M})-\gogam(\btemp_{j_{\M}})|\leq
    \left(
    \rme^{\|\btemp_{\M}\|+\|\btemp_{j_{\M}}\|}
    +
    \|\rrh\|
    \right)
    \cdot
    \|\btemp_{\M}-\btemp_{j_{\M}}\|\;\to\; 0
\]
Therefore $\gogaf(\btemp)=\lim \gogam(\btemp_{\M})$, so it follows
from \eqref{e14.2} that $B$ is the minimal point of $\gogaf(X)$.
In accordance with the Proposition \ref{tgexist} this minimum is
strict.
\end{proof}

The obtained criterion can be strengthened.

\begin{theorem}
If the sequence $\btemp_\M$ of the minimal points of $\gogam(X)$
is bounded, then it converges, and the function $\gogaf(X)$ has
strict minimum at $\btemp=\lim \btemp_{\M}$.
\end{theorem}
\begin{proof}
Since the space $\sadj$ is finite-dimensional, we can select its
converging subsequence $\btemp_{\M_{n}}\to \btemp$. Let us first
show that $\btemp=\min \gogaf(X)$. For any fixed $X$
\begin{equation}\label{e14.3}
    \gogaf(X)=\lim \gogaf_{\M_{n}}(X)\geq \;\overline{\lim} \;\gogaf_{\M_{n}}(\btemp_{\M_{n}})
\end{equation}
Now check that
\[
\;\lim \;\gogaf_{\M_{n}}(\btemp_{\M_{n}})=\gogaf(\btemp)
\]
Proceeding in a way similar to Theorem \ref{tgmexist}, we select a
sub-subsequence $\btemp_{\M_{n_j}}$ such that
$\gogaf_{\M_{n}}(\btemp_{\M_{n_j}})\to \gogaf(\btemp)$ and get the
required: the function $\gogaf(X)$ has minimum. Then, applying
Theorem \ref{tgexist} we infer that the minimum is strict and
using Theorem \ref{t14.7} we obtain that $\btemp_{\M}\to \btemp$.

\end{proof}

\section{"Temperature" theorem and its consequences}\label{shusico}

The mapping
\begin{equation}\label{ecomap}
\comap(X)=
    \int_\sepv\limits
    \rme^{\trc(X\pproj)}\;
    \pproj\, \mesn
\end{equation}
from all self-adjoint operators to positive operators in $\hh$ was
shown (proposition \ref{tgexist}) to be injective. However,
$\comap$ is not surjective: clearly, no pure product state can be
represented this way. We shall show that the image of $\comap$
contains \emph{almost} all separable density operators.

\subsection{Matching theorem}\label{shusimatch}

The decompositions of a given density operator $\rrh$ -- both
discrete and continuous are known to be non-unique. The following
theorem shows that any continuous positive decomposition of $\rrh$
can be replaced by an exponential one, which we studied before.

\begin{theorem}["Temperature" theorem]\label{ttemperature}
Let $\rrh$ be a density matrix such that it can be represented in
the form \eqref{egenrepint}
\[
\rrh\;=\;
    \int_\sepv
    w(\pproj)\cdot\pproj
    \mesn
\]
with $w$ being positive
\[
w(\pproj)>0
\]
Then there exists such self-adjoint operator $\btemp$ that:
\[
\rrh \;=\;    \int_\sepv\limits
    \rme^{\trc(\btemp\pproj)}\;
    \pproj\, \mesn
\]
\end{theorem}
\begin{proof}

Taking into account \eqref{egrad}, it suffices to prove that the
function $\gogaf(X)=\int_\sepv
    \rme^{\trc(X\pproj)}
    \mesn-\trc(X\rrh)$ has minimum. Writing
down $\rrh$ in the form \eqref{egenrepint}, we get
\begin{equation}\label{egogafposg}
\gogaf(X) \;=\; \int_\sepv\limits
    \rme^{\trc(X\pproj)}
    \mesn -
\int_\sepv\limits
    w(\pproj)\trc(X\pproj)
    \mesn
\;=\; \int_\sepv\limits\left( \rme^{\trc(X\pproj)}-
    w(\pproj)\trc(X\pproj)
    \right)
    \mesn
\end{equation}
For every fixed $\pproj\in\sepv$ introduce a function
\begin{equation}\label{efta}
g(t) \;=\; \rme^{t}- a t
\end{equation}
where
\[
a=w(\pproj)\,,\quad t=\trc(X\pproj)
\]
The set $\sepv$ is compact, and the function $w(\pproj)$ is
continuous and positive, that is why it attains its extreme values
\[
0<m\leq w(\pproj) \leq M
\]
Since $w(\pproj)$ is a probability density, we have
\begin{equation}\label{eevalm}
0<m\le 1\le M
\end{equation}
The value of $a$ in \eqref{efta} lies between $m$ and $M$. The
minimal value of $g(t)$ depends on the value of the parameter $a$
as follows:
\[
g_{\min} = a-a\ln a
\]
By elementary (for $t\leq 0$) and routine (for $t\geq 0$)
calculations we obtain
\begin{equation}\label{eevalf}
    g(t)\geq (m+M)[1-\ln(m+M)]+m|t|
\end{equation}
for any real $t$ as illustrated at this graph:
\[\includegraphics[width=70mm]{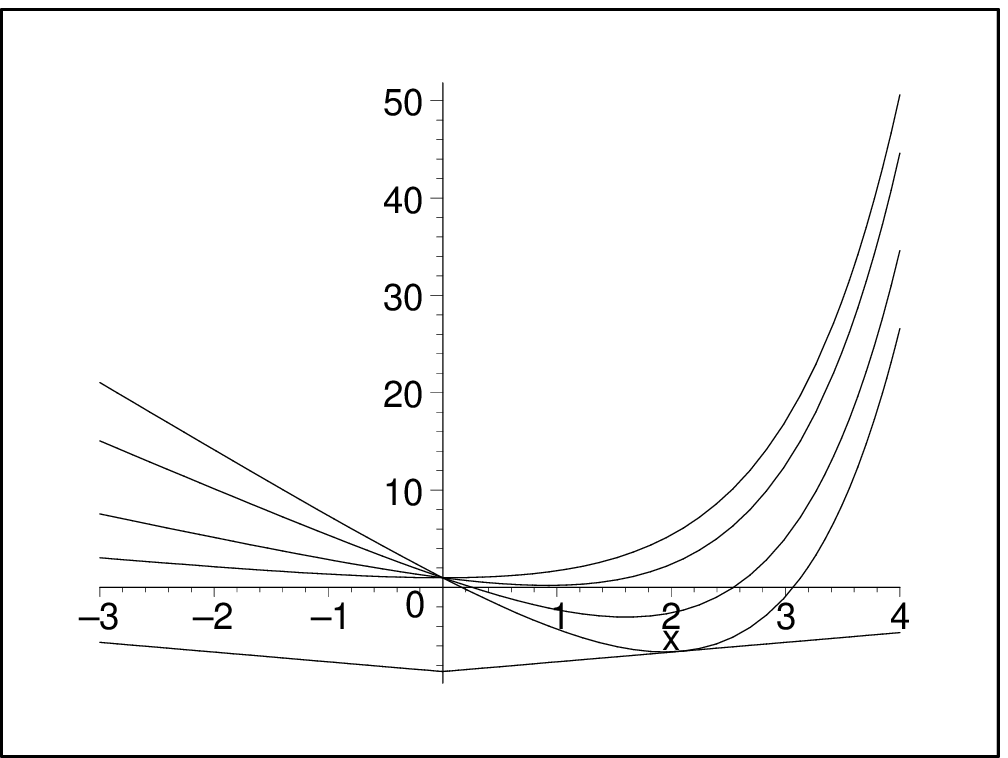}\]
Since $t=\trc(X\pproj)$ and $a=w(\pproj)$, this means
\[
\rme^{\trc(X\pproj)}- w(\pproj)\trc(X\pproj) \;\geq\;
(m+M)[1-\ln(m+M)]+m|\trc(X\pproj)|
\]
Integrating this inequality over $\sepv$, we obtain the following
evaluation:
\begin{equation}\label{eevalgogafg}
\gogaf(X) \ge (m+M)[1-\ln(m+M)] + m \int_\sepv |\trc(X\pproj)|
\mesn
\end{equation}
Consider the second summand of this expression:
\[
\nu(X) = \int_\sepv |\trc(X\pproj)| \mesn
\]
which is a seminorm. Proceeding like in Proposition \ref{tgmexist}
we see that $\nu(X)$ is non-degenerate. Since all norms in
finite-dimensional space are equivalent, $\nu(X)$ evaluates as
\[
\nu(X)\geq {\rm const}\cdot\|X\|
\]
So far, we obtain
\[
\gogaf(X) \ge (m+M)[1-\ln(m+M)] + {\rm const}\cdot \|X\|\to
+\infty
\]
Since $\gogaf(X)$ is convex, it has minimum, which is attained at
certain point $\btemp\in\sadj$.
\end{proof}

\paragraph{Remark.} It can be verified that the distribution
$w(\pproj)=\rme^{\trc(\btemp\pproj)}$
yields the maximum for
\begin{equation*}\label{evar}
    S=-\int w\ln w \mesn
\end{equation*}

\begin{theorem}[Matching theorem]\label{thusimatch}
Any $\rrh$ belonging to the interior $\intsep$ of the set of
separable states can be represented in the form \eqref{egexp}:
\[
\rrh \;=\;    \int_\sepv\limits
    \rme^{\trc(\btemp\pproj)}\;
    \pproj\, \mesn
\]
\end{theorem}
\begin{proof}
Denote by $\husrep$ the set of density operators representable as
Husimi exponentials. Let $\rrh_1,\rrh_2\in\husrep$ and form their
convex combination $\rrh=(1-t)\rrh_1+t\rrh_2$. This $\rrh$ can be
represented by continuous positive density
$w(\pproj)=(1-t)\rme^{\trc(\btemp_1\pproj)}+t\rme^{\trc(\btemp_2\pproj)}$,
therefore it follows from the "Temperature" theorem
\ref{ttemperature} that such $\btemp$ exists that
$\rrh=\int\rme^{\trc(\btemp\pproj)}\pproj\mesn$. That means,
$\husrep$ is convex together with its closure.

\medskip

Let $\pproj_0\in\sepv$ be a product pure state. Any atomic measure
on a compact set $\sepv$ can be approximated by a sequence of
strictly positive densities. Take such a sequence $w_{n}$ on
$\sepv$ such that for any continuous function $f$ on $\sepv$
\[
\lim_{n\to\infty}\int_\sepv\limits
    f(\pproj)
    w_{n}(\pproj)\;
    \, \mesn
    \;=\;
    f(\pproj_0)
\]
Consider the sequence of operators
\[
\rrh_{n} =\int_\sepv\limits
    w_{n}(\pproj)\;\pproj
    \, \mesn
\]
each belonging to $\husrep$. For any affine function $h(X)$
\[
h(\rrh_{n}) =\int_\sepv\limits
    w_{n}(\pproj)\;h(\pproj)
    \, \mesn \;\to h(\pproj_0)
\]
therefore $\rrh_{n}\to\pproj_0$. So far, the closure of $\husrep$
contains $\sepv$. The closure is shown to be convex, therefore it
contains all separable density operators.

\medskip

Now consider the "partition function"
\begin{equation}\label{epartfun}
Z(X)
    \;=\;
    \int_\sepv\limits
    \rme^{\trc(X\pproj)}
    \mesn
\end{equation}
Note that for any $V\neq 0$
\[
\rmd^2_{X} Z(V) \;=\; \left. \frac{\rmd^2 Z(X+Vt)}{\rmd t^2}
\right\vert_{t=0}
    \;=\;
    \int_\sepv\limits
    \rme^{\trc(X\pproj)}
    \bigl(\trc(V\pproj)\bigr)^2
    \mesn>0
\]
whose positivity was established in \eqref{ehesse}. However
$\rmd^2 Z=\rmd\comap$, where $\comap=\nabla Z$, see
\eqref{ecomap}. Then the mapping $\comap(X)$ is non-degenerate at
every point $X\in\sadj$. According to the inverse image theorem,
we see that $\comap$ is local diffeomorphism, that is, each point
$X$ has a neighborhood whose image under the mapping $\comap$ is
open in $\sadj$. So, we conclude that its full image
$\comap(\sadj)$ is an open subset in $\sadj$. In the meantime
\[\husrep=\comap(\sadj)\,\cap\,\{X\in\sadj\mid \trc X=1\}
\]
so, $\husrep$ is an open subset of the set of separable density
operators: $\husrep\subseteq\intsep$.

\medskip

So far, we have shown that the set $\husrep$ is an open, convex
and dense subset of the set of all separable density operators,
therefore $\husrep$ coincides with its interior:
$\husrep=\intsep$.
\end{proof}

\subsection{Husimi coordinatization}\label{scoord}

Starting from the "partition function" $Z$ defined above
\eqref{epartfun}, introduce
\begin{equation}\label{edefw}
    W=\ln Z
    \;=\;
    \ln \int_\sepv\limits
    \rme^{\trc(X\pproj)}
    \mesn
\end{equation}
and first calculate its gradient
\begin{equation*}\label{egradw}
\nabla W(X) = \frac{\nabla Z}{Z} = \frac{1}{Z} \int_\sepv\limits
    \rme^{\trc(X\pproj)}\pproj
    \mesn
\end{equation*}
\begin{proposition}\label{tnondegdw}
The quadratic form $\rmd^2 W$ is nonnegatively defined and
vanishes only on scalar operators.
\end{proposition}
\begin{proof}
Calculate the value of $\rmd^2 W$ at point $X$ on element $V$
\begin{equation*}\label{esecondw1}
    \rmd^2 W(X;V)=
    \left.\frac{\rmd^2 W(X+Vt)}{\rmd t^2}\right\vert_{t=0}=
    \left.\frac{\rmd}{\rmd t}\left(
    \frac{1}{Z} \int_\sepv\limits
    \rme^{\trc((X+Vt)\pproj)}\trc(V\pproj)
    \mesn
    \right)\right\vert_{t=0}
\end{equation*}
Denoting
\begin{equation}\label{edw}
w(P)=\dfrac{\rme^{\trc(X\pproj)}}{Z}
\end{equation}
it reads
\begin{equation}\label{esecondw}
    \rmd^2 W(X;V)=
    \int_\sepv\limits
    (\trc(V\pproj))^2
    w(P)
    \mesn
    -\left(
\int_\sepv\limits
    \trc(V\pproj)
    w(P)
    \mesn
        \right)^2
        \geq 0
\end{equation}
This expression vanishes if and only if
$\trc(V\pproj)=\mbox{const}$ for every $\pproj\in\sepv$. All such
$V$ are the scalar operators: $V=\lambda\1$.
\end{proof}
Note that for any scalar operator the function $W$ has the
property:
\begin{equation*}\label{esw}
    W(X+\lambda\1)=W(X)+\lambda
\end{equation*}
therefore its gradient
\begin{equation}\label{ehusic}
    \husic(X)=\nabla W(X)
\end{equation}
is invariant with respect to shifts along scalar operators:
\begin{equation}\label{esgradw}
   \husic(X+\lambda\1)=\husic(X)
\end{equation}
This shows that the mapping $\husic$ is not injective on $\sadj$,
but its restriction to the set $\csp$ of traceless operators from
$\sadj$ becomes injective.
\begin{theorem}\label{thusico}
The mapping \eqref{ehusic} establishes a diffeomorphism
\begin{equation}\label{ecoord}
    \husic: \csp \rightarrow \intsep
\end{equation} between
the set $\csp$ of all traceless self-adjoint operators in $\hh$
and the interior $\intsep$ of the set of all separable density
operators.
\end{theorem}
\begin{proof}
The mapping $\husic$ was shown to be invariant with respect to
shifts along scalar operators, so $\husic(\csp)=\husic(\sadj)$.
The differential $\rmd\husic=\rmd^2 W$. It follows from
Proposition \ref{tnondegdw} that $\rmd^2 W>0$ on $\csp$, therefore
$\rmd\husic$ is non-degenerate on $\csp$. So, $\husic$ is a
diffeomorphism of $\csp$ on the image $\husic(\csp)$. In turn, the
image of $\husic$ coincides with the image of the mapping
\eqref{ecomap} --- the density operators representable as \eqref{egexp}. Finally, as shown in Matching theorem
\ref{thusimatch}, this set coincides with $\intsep$ --- the
interior of the set of all separable density operators.
\end{proof}

\section*{Concluding remarks}

In this paper, the standard thermodynamical ideas were employed.
However, it was not possible to replant directly the standard
techniques. The first reason is that the analog of inverse
temperature is no longer a scalar $\beta$, it becomes self-adjoint
operator $\btemp$.  That is why the "Temperature theorem" needed
to be reproved in the new setting. One more essential difference
is the behavior of the "partition function" $Z$. Unlike classical
case, in our setting the gradient of $\ln\!Z$ is invariant
\eqref{esgradw} with respect to shifts: $X\mapsto X+\lambda\1$.

\medskip

The overall construction is general enough: we do not dwell on any
particular decomposition of the state space $\hh$ into tensor
product --- this notion was shown to be relative to measurements
\cite{paolo}. Moreover, the set of pure product states $\sepv$
could be any compact \emph{full} set of unit vectors in the state
space $\hh$. `Full' means that the value of any quadratic form in
$\hh$ is completely defined by its values on $\sepv$. This is the
central point of the method, based on multipolarization identity
\eqref{epolariz}, which was inspired by the ideas of K\^odi Husimi
\cite{husimi} to consider a special kind of functions defined on
pure states, namely, those parametrized by values of appropriate
quadratic form.

\paragraph{Acknowledgments.} The authors acknowledge the
discussion of the results offered by the participants of the
Laboratory of Quantum Information (St.Petersburg, Russia). 
A support from research grant RFFI 0706-00119 is appreciated.


\end{document}